\documentclass[journal]{IEEEtran}

\usepackage{ucs}
\usepackage[utf8x]{inputenc}
\usepackage[cmex10]{amsmath}
\usepackage{cite, amsfonts, amssymb, amsthm, bm, bbm, graphicx, relsize, multirow, booktabs, color, blindtext}
\usepackage[american]{babel}
\usepackage[T1]{fontenc}

\usepackage[keeplastbox]{flushend}

\newcommand*{\transp}{^{\mathsf{T}}}

\DeclareMathOperator*{\argmax}{\arg\,\max}
\DeclareMathOperator*{\argmin}{\arg\,\min}

\newcommand{\e}{\mathrm{e}}
\renewcommand{\i}{\mathrm{i}}
\def \SNR{\text{\relsize{-2}\sf SNR}}

\IEEEoverridecommandlockouts

\title{Radar Target Detection Aided by \\ Reconfigurable  Intelligent Surfaces}

\author{Stefano~Buzzi,~\IEEEmembership{Senior Member,~IEEE}, Emanuele~Grossi,~\IEEEmembership{Senior Member,~IEEE}, \\ Marco~Lops,~\IEEEmembership{Fellow,~IEEE}, and Luca~Venturino,~\IEEEmembership{Senior Member,~IEEE} 
	\thanks{The work of S. Buzzi, E. Grossi, and L. Venturino was supported by the research program ``Dipartimenti di Eccellenza 2018--2022'' sponsored by the Italian Ministry of Education, University, and Research (MIUR).}
	\thanks{Stefano~Buzzi, Emanuele~Grossi, and Luca~Venturino are with the Department of Electrical and Information Engineering (DIEI), University of Cassino and Southern Lazio,  03043 Cassino, Italy, and with Consorzio Nazionale Interuniversitario per le Telecomunicazioni, 43124 Parma, Italy (e-mail: buzzi@unicas.it, e.grossi@unicas.it, l.venturino@unicas.it).}
	\thanks{Marco~Lops is with the Department of Electrical and Information Technology (DIETI), University of Naples \emph{Federico II}, 80138 Naples, Italy, and with Consorzio Nazionale Interuniversitario per le Telecomunicazioni, 43124 Parma, Italy (e-mail: lops@unina.it).}
}

\begin{document}
\bstctlcite{BSTcontrol}

This work has been submitted to the \textit{IEEE Signal Processing Letters} for possible publication. Copyright may be transferred without  notice,  after  which  this  version  may  no  longer  beaccessible.

\maketitle

\begin{abstract}
 In this work, we consider the target detection problem in a sensing architecture where the radar is aided by a reconfigurable intelligent surface (RIS), that can be modeled as an array of sub-wavelength small reflective elements capable of imposing a tunable phase shift to the impinging waves and, ultimately, of providing the radar with an additional echo of the target. A theoretical analysis is carried out for closely- and widely-spaced (with respect to the target) radar and RIS and for different beampattern configurations, and some examples are provided to show that large gains can be achieved by the considered detection architecture.
\end{abstract}

\section{Introduction}

One of the most striking technological innovations of the recent past in the field of radio communications is represented by metasurfaces~\cite{Tsilipakos-2020}, and in particular by reconfigurable intelligent surfaces (RISs)~\cite{RIS_IEEE_ACCESS,wu2019towards,Larsson-ThreeMyth2020}. Traditionally, wireless communication and radar systems have been realized based on a proper design and optimization of the transmitter and of the receiver, assuming that no action could be taken to improve the channel propagation characteristics. This paradigm has been lately challenged by the introduction of RISs, which are man-made thin surfaces whose electromagnetic response can be electronically controlled. RISs are nearly passive devices, with very low energy consumption, which have the capability of tuning the phase, amplitude, frequency, and polarization of reflected impinging wavefronts~\cite{RIS_IEEE_ACCESS}. As such, they introduce further degrees of freedom to be exploited for system optimization and allow shaping the wireless channel impulse response. They can be mounted outdoors on building facades or in indoor environments on the ceiling or on walls. 

RISs have attracted a lot of interest in the recent past, and several studies have shown their usefulness for wireless communications. E.g., \cite{Zap_RIS}  has considered the downlink multiuser communication in a single-cell network with multiple antennas at the base station, and the RIS phase offsets have been optimized  to increase the system energy efficiency; in~\cite{mursia2020risma}, instead, the problem of massive access for IoT devices has been considered, and it is shown that the RIS provides remarkable gains to the system sum-rate. In~\cite{Huan19}, the authors have considered an indoor placement and have configured the RIS phases through a deep neural network. The use of a RIS in the context of millimeter-wave ultra massive MIMO systems has been investigated in~\cite{2021-Tulino}. Finally, the RIS-assisted localization has been investigated in~\cite{wymeersch2020radio, He-2020, elzanaty2020reconfigurable}.

All of these works have focused on the optimization of wireless communication systems, for either data exchange or localization (where the network is aware of the existence of a mobile device and can exploit the signals actively transmitted by the collaborating device). The possible benefits that a RIS could bring to a radar system in enhancing its detection capabilities are, to be best of the authors' knowledge, a still unexplored topic, and this letter offers a first contribution aimed at filling this gap. We consider a scenario where the radar can transmit (or receive) through two separate beams, one pointing in the inspected direction and one pointing towards the RIS, which is aimed at focusing the impinging wavefront towards the prospective target during the transmission phase or towards the radar during the reception phase. In this framework, we provide key conditions for system design, that relate the signal bandwidth, the carrier frequency, and the sizes and mutual positions of radar, RIS, and prospective target; we inspect two relevant scenarios, where the radar and the RIS are closely- and widely-spaced, and we show that in both scenarios the RIS phases can be properly adjusted so as to \emph{align} the echoes reaching the radar; we analyze two examples of realistic applications, and we show that the RIS can bring substantial improvements to the received signal-to-noise ratio (SNR); finally, we provide a detailed discussion, where we come up with relevant insights for system design, draw the conclusions, and outline future developments.

\section{System model}

\begin{figure}[t]
 \centering
 \centerline{\includegraphics[width=0.93\columnwidth]{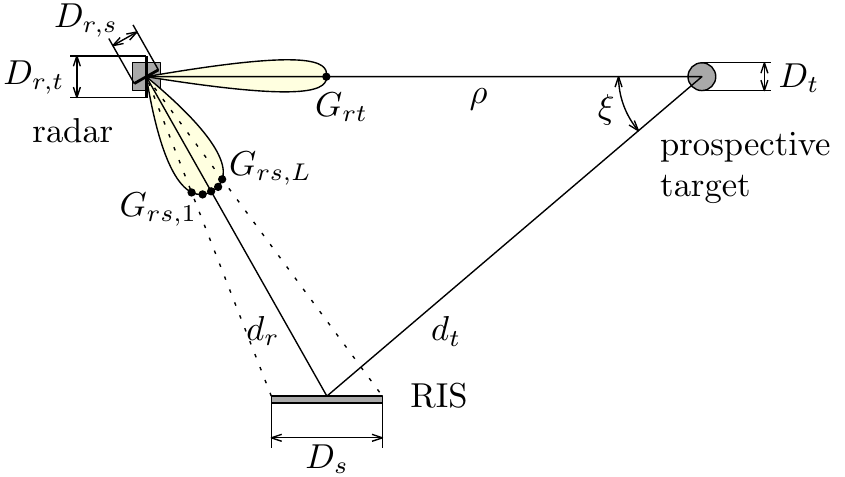}}
 \caption{Considered architecture composed of a radar aided by an RIS.} \label{fig_1} \vspace{-2pt}
\end{figure}

Consider a target detection problem where the radar is assisted by an RIS, as shown in Fig.~\ref{fig_1}, and denote by $P_r$ the radar transmit power, $\lambda$ the carrier wavelength, $L$  the number of sub-$\lambda$-sized surface element of the RIS, $\rho$, $d_{r,\ell}$ ($d_r$), and $d_{t,\ell}$ ($d_t$) the distances between the radar and the target, the $\ell$-th element (the center) of the RIS and the radar, and the $\ell$-th element (the center) of the RIS and the target, respectively, and $G_{rt}$ and $G_{rs,\ell}$ the gains of the radar beams towards the target and the $\ell$-th element of the RIS, respectively. We assume that the radar waveform is narrowband, so that the delays of the target echo reaching the elements of the RIS and of the target echo reaching the radar are not resolvable, i.e., $\max\{D_{rt}, D_{rs}, D_s\} \ll c/W$, where $D_{rt}$, $D_{rs}$, and $D_s$ are the size of the radar antenna pointing towards the target, of the radar antenna pointing towards the RIS,\footnote{The radar may be equipped either with a single antenna capable of forming two beams or with two distinct antennas.} and of the RIS, respectively, $c$ is the speed of light, and $W$ is the radar signal bandwidth. Furthermore, we assume $\min\{D_{rt},D_{rs}\}\gg\lambda$, i.e., that the radar antennas are directive, and $D_s\gg \lambda$. Finally, we assume that the impinging wavefield can be approximated as a plane wave in the paths between radar and target, target and RIS, and radar and each element of the RIS; namely, we assume that the destinations in the aforementioned hops (in both directions) are in the far-field~\cite{book-Stutzman02, Balanis_2012}, i.e.,
\begin{equation}
 \begin{cases}
 \rho \geq 2 \max\{D_{rt}^2, D_t^2\}/\lambda\\
 \min \{d_{t,\ell} \}_{\ell=1}^L\geq 2 \max \{D_t^2, D_s^2\}/\lambda\\
 \min \{d_{r,\ell} \}_{\ell=1}^L \geq 2D_{rs}^2/\lambda
 \end{cases}\label{far_field}
\end{equation}
where $D_t$ is the size of the target. Observe that the whole RIS and the radar may not be in the far-field of each other.

In this framework, we analyze the following cases.
\begin{itemize}
 \item \emph{Case~a}: the radar has one transmit beam pointing towards the target and two receive beams pointing towards the target and the RIS.
 \item \emph{Case~b}: the radar has two transmit beams pointing towards the target and the RIS and one receive beam pointing towards the target; the delays corresponding to the direct and indirect echoes are resolvable,\footnote{Notice that this case can be enforced also when the delays are not resolvable by transmitting orthogonal waveforms from the two beams.} namely, $d_t+ d_r - \rho \geq c/W$.
\item \emph{Case~c}: same as Case~b, but the delays are not resolvable, i.e., $d_t+ d_r - \rho \ll c/W$.
\end{itemize}
After matched-filtering with the transmit waveform and sampling, if a target is present in the resolution cell under test,\footnote{Here we make the standard assumption of neglecting the system losses caused by the possible mismatch between the sampling instant and/or pointing angle and the true delay and/or angle of the target (beam-shape and straddling losses~\cite{Skolnik_2001}); we also neglect the signal components due to the sidelobes of the beampatterns and, for Case b, of the autocorrelation function of the transmit waveform.} the received discretized signal can be written as
\begin{subequations}
\begin{gather}
 \begin{pmatrix} x_1\\ x_2 \end{pmatrix} = \begin{pmatrix} \alpha \sqrt{\sigma}\e^{\i \beta} \\  \sum_{\ell=1}^L \alpha_{sr,\ell}\sqrt{\sigma_s}  \e^{\i(\psi_{t,\ell}+\phi_\ell+\psi_{r,\ell})} \end{pmatrix} + \begin{pmatrix} w_1\\ w_2 \end{pmatrix}\label{r_case_1}\\
  \begin{pmatrix} x_1\\ x_2 \end{pmatrix} = \begin{pmatrix} \alpha \sqrt{\sigma \epsilon} \e^{\i \beta} \\  \sum_{\ell=1}^L \alpha_{st,\ell}\sqrt{\sigma_s(1-\epsilon)}  \e^{\i(\psi_{r,\ell}+\phi_\ell+\psi_{t,\ell})}\end{pmatrix} + \begin{pmatrix} w_1\\ w_2 \end{pmatrix}\label{r_case_2}\\
x_1 = \alpha \sqrt{\sigma \epsilon} \e^{\i \beta}+ \sum_{\ell=1}^L \alpha_{st,\ell}\sqrt{\sigma_s(1-\epsilon)}  \e^{\i(\psi_{r,\ell}+\phi_\ell+\psi_{t,\ell})}  + w_1  \label{r_case_3}
\end{gather} \label{r_all_cases}%
\end{subequations}
for Case~a,~b, and~c, respectively, where: $\sigma$ and $\sigma_s$ are the (unknown) target radar cross-sections (RCSs) observed from the radar and from the RIS, respectively; $\beta$ is the (unknown) phase of the radar-target channel; $\{\psi_{r,\ell}\}_{\ell=1}^L$ are the phases of the RIS-radar channel; $\{\phi_\ell\}_{\ell=1}^L$ are the (adjustable)  RIS phases; $\{\psi_{t,\ell}\}_{\ell=1}^L$ are the (unknown) phases of the RIS-target channel; $w_1$ and $w_2$ are the noise components, modeled as independent and identically distributed complex circularly symmetric Gaussian random variables with variance $P_w$; $\epsilon\in[0,1]$ is the fraction of transmit power allocated to the direction of the target for Cases~b and~c; and, from the radar equation,
\begin{gather}
\alpha = \sqrt{\frac{P_r G_{rt}^2  \lambda^2 }{(4\pi)^3 \rho^4}}, \quad  \alpha_{sr,\ell} = \sqrt{\frac{P_r G_{rt} G_{rs,\ell} \lambda^2 S_{sr,\ell}}{(4\pi)^4 \rho^2 d_t^2 d_{r,\ell}^2}}\\
 \alpha_{st,\ell} = \sqrt{\frac{P_r G_{rt} G_{rs,\ell} \lambda^2 S_{st,\ell}}{(4\pi)^4 \rho^2 d_t^2 d_{r,\ell}^2}}
\end{gather}
with $S_{sr,\ell}$ and $S_{st,\ell}$ denoting the (bistatic) RCSs of the $\ell$-th reflecting element of the RIS in the target-RIS-radar and radar-RIS-target  paths, respectively.

The problem is to optimize the architecture in Fig.~\ref{fig_1} so as to improve the detection performance of the radar by exploiting the additional target echo available from the RIS.

\section{System design} \label{system_opt_sec}

Depending on the size of the target, the antenna wavelength, and the mutual distances among radar, target, and RIS, we have two relevant situations. Namely, if $\xi \ll \lambda/ D_t $, where $\xi$ is the angle formed by the line segment linking the target and the radar and the line segment linking the target and the RIS (cfr. Fig.~\ref{fig_1}), then radar and RIS are seen as \emph{closely-spaced} (or \emph{co-located}) by the target~\cite{Fishler_2006}, since they lie in the same angular beam of the target scattering (that is proportional to $\lambda/D_t$). On the opposite extreme, we have the situation where radar and RIS are \emph{widely-spaced} with respect to the target, i.e., $\xi \geq \lambda/ D_t $. In the following, we address these two situations.

\subsection{Closely-spaced radar and RIS}

In this case, we have that $\sigma_s=\sigma$ and $\psi_{t,\ell}=\psi_{t,\ell}'+\beta$, for $\ell=1,\ldots,L$, where $\psi_{t,\ell}'$ is known.\footnote{Radar and RIS are struck by a plane wave from the target, so that the difference between the phase delay at the radar, $\beta$, and the phase delay at the $\ell$-th element of the RIS, $\psi_{t,\ell}$, is uniquely determined by their mutual positions with respect to the target.} Since $\psi_{r,\ell}$ can be estimated, the best performance is achieved when the RIS phases are chosen as $\phi_\ell=-\psi_{t,\ell}'-\psi_{r,\ell}$, so that all the signal terms are \emph{phase-aligned}, and~\eqref{r_all_cases} become
\begin{subequations}
\begin{align}
\begin{pmatrix} x_1 \\ x_2 \end{pmatrix} &= \sqrt{\sigma}\e^{\i \beta} \begin{pmatrix} \alpha \\ \alpha_{sr}\end{pmatrix}+ \begin{pmatrix} w_1 \\ w_2 \end{pmatrix}\\
\begin{pmatrix} x_1 \\ x_2 \end{pmatrix} &= \sqrt{\sigma}\e^{\i \beta} \begin{pmatrix} \alpha\sqrt{\epsilon}\\ \alpha_{st} \sqrt{1-\epsilon} \end{pmatrix}+ \begin{pmatrix} w_1 \\ w_2 \end{pmatrix}\\
 x_1 &= \sqrt{\sigma}\e^{\i \beta} (\alpha\sqrt{\epsilon}  + \alpha_{st} \sqrt{1-\epsilon}) + w_1
 \end{align}%
\end{subequations}
where $\alpha_{sr} =\sum_{\ell=1}^L \alpha_{sr,\ell}$ and $\alpha_{st} = \sum_{\ell=1}^L \alpha_{st,\ell}$. The generalized likelihood ratio tests (GLRTs)~\cite{Van_Trees_1} with respect to the unknown complex parameter $\sqrt{\sigma}\e^{\i\beta}$ are
\begin{equation}
\begin{cases}
 \frac{\left| \alpha x_1+ \alpha_{sr} x_2 \right|^2}{(\alpha^2+ \alpha_{sr}^2)P_w} \gtrless \gamma,& \text{for Case~a}\\
  \frac{\left| \alpha \sqrt{\epsilon} x_1+ \alpha_{st} \sqrt{1-\epsilon} x_2 \right|^2}{(\alpha^2\epsilon + \alpha_{st}^2(1-\epsilon))P_w} \gtrless \gamma,& \text{for Case~b}\\
\frac{|x_1|^2}{P_w}  \gtrless \gamma,& \text{for Case~c}
 \end{cases}
\end{equation}
where $\gamma>0$ is the detection threshold.

Denoting $\bar \sigma= \mathbbm E[\sigma]$, we have that the SNR is
\begin{subequations}
\begin{align}
 \SNR_a&= \frac{\bar \sigma(\alpha^2+ \alpha_{sr}^2)}{P_w} =\SNR_0 (1+K_{sr})\\
 \SNR_b&=\frac{\bar\sigma \alpha^2 \epsilon +\alpha_{st}^2 (1-\epsilon)}{P_w} \leq  \SNR_0 \max\{1, K_{st}\}\\
 \SNR_c&=\frac{\bar\sigma (\alpha \sqrt{\epsilon} +\alpha_{st} \sqrt{1-\epsilon})^2}{P_w} \leq  \SNR_0 (1+K_{st})
\end{align}\label{SNR_closely_spaced}%
\end{subequations}
where $ \SNR_0 = \alpha^2 \bar \sigma/P_w= P_r G_{rt}^2\lambda^2 \bar\sigma/\bigl((4\pi)^3 \rho^4 P_w\bigr)$ is the SNR when the RIS is absent, and
\begin{align}
 K_{sr} &= \frac{\alpha_{sr}^2}{\alpha^2}= \frac{\rho^2}{4\pi d_t^2 G_{rt}} \left( \sum_{\ell=1}^L \frac{\sqrt{G_{rs,\ell}S_{sr,\ell}}}{d_{r,\ell}} \right)^2\\ 
 K_{st} &= \frac{\alpha_{st}^2}{\alpha^2}=  \frac{\rho^2}{4\pi d_t^2G_{rt}} \left( \sum_{\ell=1}^L \frac{\sqrt{G_{rs,\ell} S_{st,\ell}}}{d_{r,\ell}} \right)^2
\end{align}
are the power gains of the indirect echoes with respect to the direct one.
Observe that the upper bounds on $\SNR_b$ and $\SNR_c$ are achieved by properly splitting the transmit power over the two available beams, i.e., by using $\epsilon_b^*=\mathbbm 1_{\{K_{st}\leq 1\}}$, and $\epsilon_c^*= 1/(1+K_{st})$, where $\mathbbm 1_{\mathcal A}=1$, if the condition $\mathcal A$ holds true, and $\mathbbm 1_{\mathcal A}=0$, otherwise.

The probability of false alarm is $P_\text{fa}=  \e^{-\gamma}$, while the probability of detection, $P_\text{d}$, can be found once the distribution of $\sigma$ is given, and it admits the well-known, closed-form expressions for the Marcum's non-fluctuating case and for the Swerling's models~\cite{Richards_2005}.

\subsection{Widely-spaced radar and RIS}

In this case, denoting $\beta_s=\psi_{t,1}$, we have that $\psi_{t,\ell}=\psi_{t,\ell}''+\beta_s$, where $\psi_{t,\ell}''$ is known.\footnote{The RIS is truck by a plane wave, and the difference between the phase delay on the first element, $\psi_{t,1}=\beta_s$, and the phase delay on the $\ell$-th element, $\psi_{t,\ell}$, is uniquely determined by their mutual positions with respect to the target.} Again, the RIS phases can be chosen as $\phi_\ell=-\psi_{t,\ell}''-\psi_{r,\ell}$, so that~\eqref{r_all_cases} become
\begin{subequations}
\begin{align}
 \begin{pmatrix}  x_1 \\ x_2 \end{pmatrix}  & = \begin{pmatrix}  \sqrt{\sigma}\e^{\i \beta} \alpha \\  \sqrt{\sigma_s}\e^{\i \beta_s} \alpha_{sr} \end{pmatrix} + \begin{pmatrix}  w_1 \\ w_2 \end{pmatrix}\\
 \begin{pmatrix}  x_1 \\ x_2 \end{pmatrix}  & = \begin{pmatrix}  \sqrt{\sigma\epsilon}\e^{\i \beta} \alpha \\   \sqrt{\sigma_s (1-\epsilon)}\e^{\i \beta_s} \alpha_{st} \end{pmatrix} + \begin{pmatrix}  w_1 \\ w_2 \end{pmatrix}\\
x_1& = \sqrt{\sigma\epsilon}\e^{\i \beta} \alpha +  \sqrt{\sigma_s (1-\epsilon)}\e^{\i \beta_s} \alpha_{st}+ w_1 
\end{align}%
\end{subequations}
where $\sqrt{\sigma} \e^{\i\beta}$ and $\sqrt{\sigma_s} \e^{\i\beta_s}$ can be modeled as independent random variables, and the GLRTs are
\begin{equation}
\begin{cases}
 \frac{|x_1|^2 + |x_2|^2}{P_w} \gtrless \gamma ,& \text{for Cases~a and~b}\\
 \frac{|x_1|^2}{P_w} \gtrless \gamma,& \text{for Case~c.}
\end{cases}
\end{equation}

Denoting $\bar \sigma_s =\mathbb E[\sigma_s]$, the SNRs for the two observations in Cases~a and~b are
\begin{equation}
\begin{cases}
 \SNR_{a,1}= \SNR_0,\\
  \SNR_{a,2}= \SNR_0 K_{sr} \bar \sigma_s/\bar \sigma,
 \end{cases}
 \begin{cases}
  \SNR_{b,1}= \epsilon \SNR_0, \\
  \SNR_{b,2}= (1-\epsilon)\SNR_0 K_{st} \bar \sigma_s/\bar \sigma.
\end{cases}\label{SNR_ab_widely_spaced}
\end{equation}
For Case~c, assuming $\beta$ and $\beta_s$ uniformly distributed over $[0,2\pi)$, we have
\begin{equation}
 \SNR_c=\bigl(\epsilon \bar\sigma \alpha^2 + (1-\epsilon) \bar\sigma_s \alpha_{st}^2\bigr)/P_w.
\end{equation}
Notice that $\epsilon$ in Cases~b and~c cannot be directly optimized, since  the average RCSs of the target are in general unknown. If $\bar \sigma$ and $\bar \sigma_s$ are assumed to lie in $[\bar \sigma_\text{min}, \bar \sigma_\text{max}]$ and $[\bar \sigma_{s,\text{min}}, \bar \sigma_{s,\text{max}}]$, respectively, the choice that maximizes the worst-case $P_d$ is obtained for Case~c by maximizing the worst-case SNR: specifically, we have  $\epsilon_c^*=\mathbbm 1_{\{K_{st} \bar \sigma_{s,\text{min}} / \bar \sigma_\text{min} \leq 1 \}}$ that gives
\begin{equation}
 \SNR_c \!= \!\SNR_0 \bigl(\mathbbm 1_{\{K_{st} \bar \sigma_{s,\text{min}} / \bar \sigma_\text{min} \leq 1  \}} +\mathbbm 1_{\{K_{st} \bar \sigma_{s,\text{min}} / \bar \sigma_\text{min} \geq 1 \}} K_{st} \bar \sigma_s/ \bar \sigma\bigr). \label{SNR_c_widely}
\end{equation}
For Case~b, instead, $\epsilon_b^*$ depends on the target fluctuation model.

The probability of false alarm is $P_\text{fa}= \e^{-\gamma} (1+\gamma)$ for Cases~a and~b and $P_{fa}=\e^{-\gamma}$ for Case~c~\cite{Richards_2005}, while, again, the probability of detection can be found once the distribution of $\sigma$ and $\sigma_s$ is given.

\vspace{-5pt}
\section{Examples}\label{num_res_sec}

\begin{figure}[t]
 \centering
 \centerline{\includegraphics[width=\columnwidth]{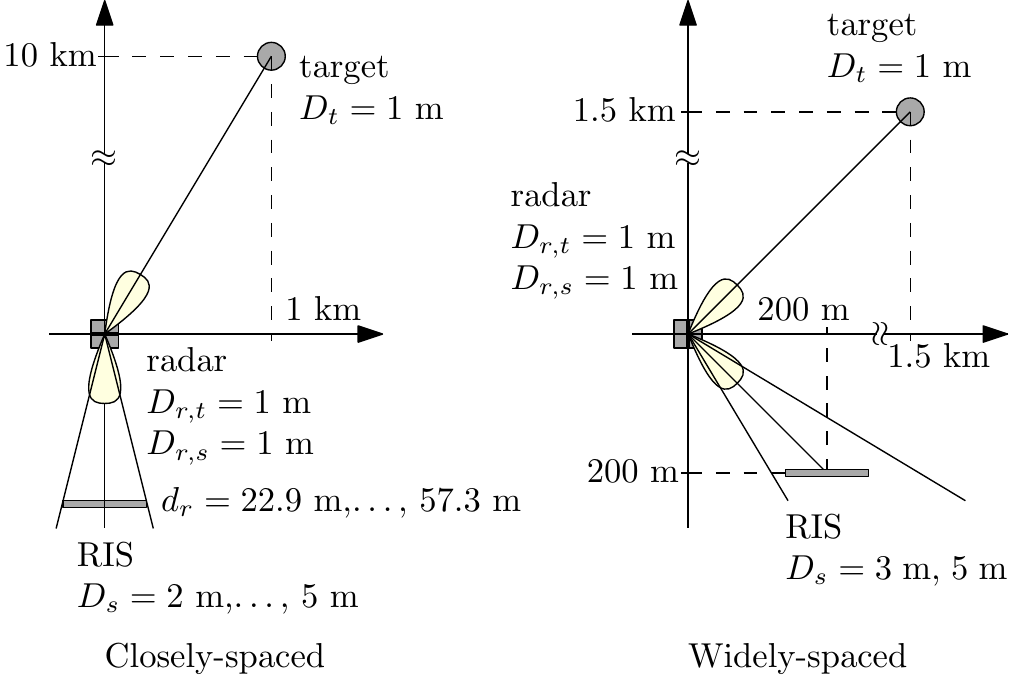}}
\caption{System geometry in the closely- and widely-spaced scenarios.\label{fig_2}}
\end{figure}

We consider a radar operating at 3~GHz equipped with two $1\times1$~m uniform square arrays with a $\lambda/2$ element spacing and a cosine element pattern in both azimuth and elevation,\footnote{This is done here just to simplify the exposition, but cheaper solutions may be devised, since the antenna pointing towards the RIS is steady.}  and we examine two bandwidths, 1 and 10~MHz. The RIS is a square surface with inter-element spacing $\lambda/2$; the RCS of each element is modeled as $S_{sr,\ell}=S_{st,\ell}=\pi (\lambda/2)^2 \cos\theta_t \cos \omega_t \cos\theta_{r,\ell} \cos \omega_{r,\ell}$, where $(\theta_t,\omega_t)$ and $(\theta_{r,\ell},\omega_{r,\ell})$ are the angles of incidence (azimuth, elevation) of the wave on the $\ell$-th element of the RIS from the target and the radar, respectively.\footnote{This is a simple yet realistic model, where the RCS of the elements is the product of an effective aperture and a gain, with a scan loss in the form of some power of the cosine~\cite{Ellingson2019ris, DiRenzo2021-modeling-pathloss}. It also implies that, at broadside direction, the physical area of the RIS is equal to the sum of the effective apertures of the elements, and the RCS of the RIS becomes the (monostatic) RCS of a flat plate of the same size. Notice that the pair $(\theta_t,\omega_t)$ does not depend on $\ell$, since target and RIS are in the far-field.} The geometry of the system is depicted in Fig.~\ref{fig_2}, with the RIS parallel to the $x$-$z$ plane.

In the closely-spaced scenario, the radar antennas lie on the $x$-$z$ plane, and different RIS sizes are tested, ranging from 2 to 5~m; specifically, $d_r$ increases with the RIS size in such a way that the area covered by the 3-dB beamwidth of the radar antenna is equal to the RIS surface area. In the configuration with two transmit and one receive beam, Case~b always holds when $W=10$~MHz, while Case~c holds for\footnote{The radar range-cell size is $c/(2W)$, and we used half of this size to check the condition that direct and indirect paths are not resolvable.}  $D_s\leq3$~m when $W=1$~MHz. The SNR gain, measured by the ratio $\SNR_{a,b,c}/\SNR_0$, is reported in Table~\ref{closely_spaced_table}: it is seen by inspection that a large performance improvement can be achieved in all operating conditions.

\begin{table}\caption{SNR gain [dB] in the closely-spaced scenario \vspace{-5pt}\label{closely_spaced_table}}
\begin{center}
\begin{tabular}{lccccccc}
\toprule
 $D_s$ [m] & 2 & 2.5 & 3 & 3.5 & 4 & 4.5 & 5  \\
 $d_r$ [m] & 22.9 & 28.6 & 34.4 & 40.1 & 45.8 & 51.5 & 57.3 \\
\midrule
Case~a& 4.78 & 6.17 & 7.41 & 8.53  & 9.54  &  10.5  & 11.3 \\
Case~b (10 MHz) & 3.03 & 4.96 & 6.54  & 7.88 &  9.03  & 10.0  &  11.0 \\
Case~c (1 MHz) & 4.78 & 6.17 & 7.41  & --- & --- & --- & --- \\
\bottomrule
\end{tabular}
\end{center}
\end{table}

\begin{figure}[t]
 \centering
 \vspace{-12pt}\centerline{\includegraphics[width=1.1\columnwidth]{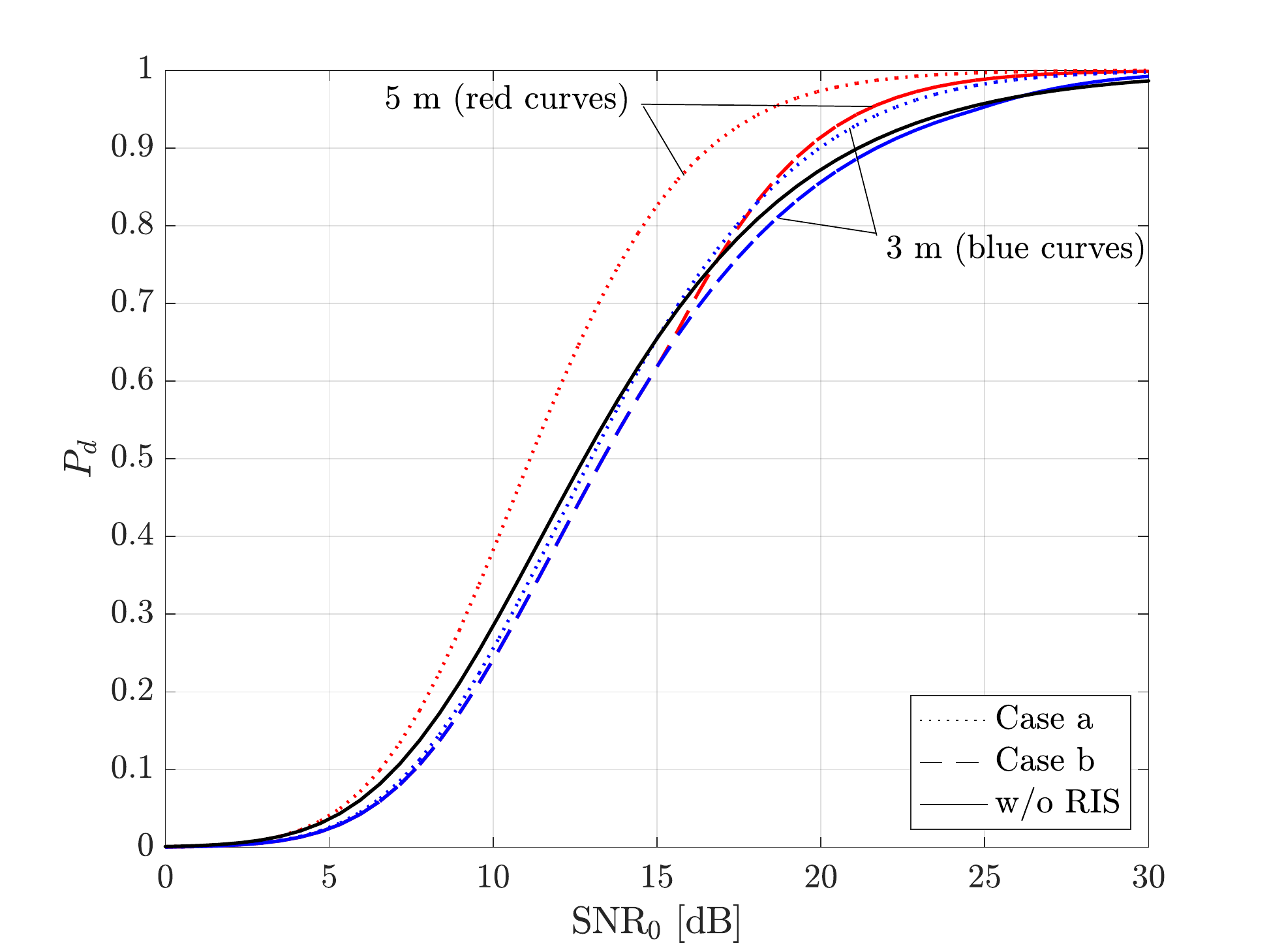}}
 \vspace{-7pt}
\caption{Probability of detection in the widely-spaced scenario; $P_\text{fa}=10^{-6}$.} \label{fig_3}
\end{figure}

In the widely-spaced scenario, the radar antennas lie on the $y$-$z$ plane, while the RIS has a side length of either 3 or 5~m. In the configuration with two transmit and one receive beam, Case~b always holds for both bandwidths. An exponential fluctuation model is considered for the RCS of the target (a closed-form expression for $P_d$ is therefore available for all cases), and the parameter $\epsilon$ for Case~b is selected so as to maximize the worst-case $P_\text{d}$ when $\bar \sigma_{s,\text{min}}=\bar \sigma_\text{min}$. Fig.~\ref{fig_3} shows $P_\text{d}$ as a function of $\SNR_0$ when $P_\text{fa}=10^{-6}$ and $\bar\sigma_s=\bar\sigma$; the case without RIS has been included for the sake of comparison. It can be seen that a performance improvement is obtained with the RIS, and that Case~a always outperforms Case~b: indeed, since $K_{sr}= K_{st}$, the SNR's in Case~a are larger than those in Case~b.

\section{Discussion}\label{discussion_sec}

Some useful insights on the performance improvement granted by the RIS can be gained, if the following approximations are made. Assume that $\theta_{r,\ell}\approx \theta_r$, $\omega_{r,\ell}\approx \omega_r$, $d_{r,\ell}\approx d_r$, and $G_{rs,\ell}\approx  4\pi d_r^2 / A_{rs}$, where $A_{rs}$ is the surface area covered by the 3-dB beamwidth of the radar antenna at a distance equal to $d_r$, if the $\ell$-th element of the RIS falls in the 3-dB beamwidth, and $G_{rs,\ell}\approx0$, otherwise. Let $A_{sr}=L (\lambda/2)^2 \cos\theta_r \cos \omega_r$ be the effective area of the RIS seen from the radar, and $G_{st}=L \pi \cos\theta_t \cos \omega_t$ be the gain of the RIS (seen as an aperture antenna) towards the target. Then, $S_{st,\ell} = S_{sr,\ell} A_{sr} \approx G_{st} /L^2$, and
\begin{equation}
K_{sr}=K_{st}\approx  \frac{\rho^2 G_{st}}{d_t^2 G_{rt}} \min\left\{\frac{A_{sr}}{A_{rs}}, \frac{A_{rs}}{A_{sr}}\right\}
\end{equation}
since the summation over $\ell$ encompasses about $L \min\{1,A_{rs}/A_{sr}\}$ terms. This shows that the RIS should be large and close enough so as to \emph{fill} the area covered by the radar beam (as done in the closely-spaced example of Sec.~\ref{num_res_sec}), i.e., $A_{sr}\approx A_{rs}$: in this case, $K_{sr}=K_{st}\approx \rho^2G_{st}/(d_t^2 G_{rt})$, which can be quite large for a small, low-cost radar and a large RIS. Basically, the radar beam pointing towards the RIS acts as a feed antenna, that sends/receives the signal via a large reconfigurable surface capable of electronically tunable beamforming. This can also be seen by noticing that the SNR corresponding to the indirect echo is
\begin{equation}
\frac{\bar \sigma \alpha_{st}^2}{P_w} =\frac{\bar \sigma \alpha_{sr}^2}{P_w} \approx \frac{P_r G_{rt} G_{st} \lambda^2 \bar \sigma}{(4\pi)^3 \rho^2 d_t^2 P_w}
\end{equation}
that is just the radar equation, where the transmit or receive radar gain has been replaced by the RIS gain. If $G_{st}$ is much larger than $G_{rt}$, one may think to use only the beam pointing towards the RIS for both transmitting and receiving.

Finally, it is worthwhile noticing that, in the widely-spaced scenario, the proposed system realizes a low-cost bistatic radar, where a two-fold diversity is available in Cases~a and~b, thanks to the two observations with different aspect angles; this may also improve the estimation capabilities of the radar.

To conclude, the goal of this letter was to show that RISs can play a crucial role also in radar applications, and to unveil first fundamental trade-offs and main issues. Accordingly, a basic and simple setting was considered. Further research is needed to ascertain the beneficial effects of the RIS in more complex and challenging scenarios involving for instance the use of MIMO radars and of multiple distributed RISs, that can be simultaneously used in the transmit and receive phase.

\appendix

\setcounter{figure}{0}
\renewcommand{\thefigure}{A\arabic{figure}}

On explicit request of one of the anonymous reviewers, we report here the detailed derivations of the main results and equations.

\renewcommand{\theequation}{A.\arabic{equation}}

\subsection{RIS-target channel phases in Footnote~4}

\begin{figure}[t]
 \centering
 \centerline{\includegraphics[width=\columnwidth]{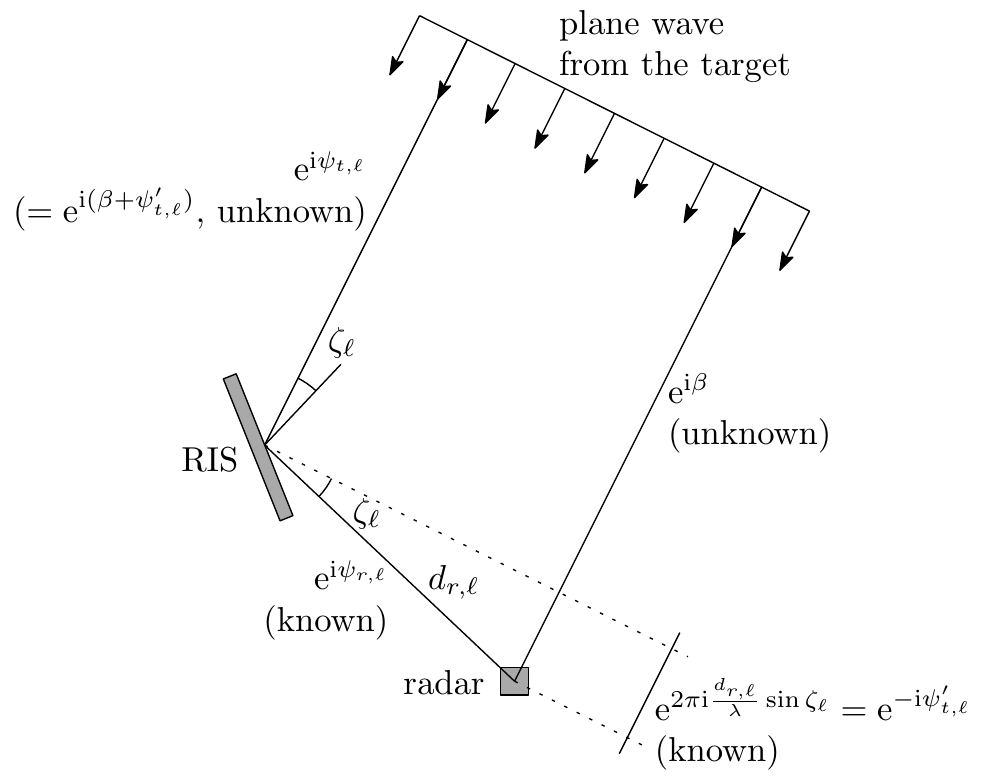}}
\caption{Phases of the RIS-target channel in the closely-spaced scenario.} \label{fig_4}
\end{figure}

Radar and RIS \emph{see} the target from the same angle, and the phases of the target-RIS channel, $\{\psi_{t,\ell}\}_{\ell=1}^L$, can be expressed as $\psi_{t,\ell}'+\beta$, where $\psi_{t,\ell}'$ is known and depends only on the mutual position of the radar and the $\ell$-th element of the RIS with respect to the target, while $\beta$ is the phase of the target-radar channel, as it can also be seen from Fig.~\ref{fig_4}. In this case, we have the same radar cross-section (RCS) of the target, $\sqrt{\sigma} \e^{\i\beta}$, in both echoes.

\subsection{GLRT's in Eqs.~(6)}

The observation vector $\bm x=(x_1 \; x_2)\transp \in \mathbb C^2$ in Eq.~(5a) is
\begin{equation}
 \bm x= \sqrt{\sigma}\e^{\i \beta} \begin{pmatrix} \alpha \\ \alpha_{sr}\end{pmatrix}+ \begin{pmatrix} w_1 \\ w_2 \end{pmatrix} = A \bm \alpha + \bm w\label{obs_model}
\end{equation}
where $\bm \alpha=(\alpha \; \alpha_{sr})\transp \in \mathbb R^2$, $A=\sqrt{\sigma} \e^{\i \beta}\in\mathbb C$ is the target response, modeled as an unknown deterministic parameter, and $\bm w=(w_1 \; w_2)\transp \in \mathbb C^2$ is a complexy circularly symmetric Gaussian random vector with covariance matrix $P_w \bm I$. Therefore, the density of $\bm x$ is
\begin{equation}
 \begin{cases}
  f_1(\bm x; A) =\frac{1}{(\pi  P_w)^2} \e^{-\frac{1}{ P_w} \Vert \bm x - A \bm \alpha\Vert^2}, & \text{if the target is present}\\
  f_0(\bm x) =\frac{1}{(\pi  P_w)^2} \e^{-\frac{1}{ P_w} \Vert \bm x \Vert^2}, & \text{otherwise}
 \end{cases}
\end{equation}
and the maximum likelihood (ML) estimate of $A$ under the ``target presence'' hypothesis is
\begin{align}
 \hat A_\text{ML} &= \argmax_{A \in \mathbb C}  f_1(\bm x; A) \notag\\
 & = \argmin_{A \in \mathbb C}  \Vert \bm x - A \bm \alpha\Vert^2 \notag\\
 &= \frac{\bm \alpha\transp \bm x}{\Vert \bm\alpha \Vert^2}.
\end{align}
Thus, the GLRT is
\begin{align}
 \frac{f_1(\bm x; \hat A_\text{ML})}{f_0(\bm x)}& = \e^{\frac{1}{ P_w} \left( \Vert \bm x\Vert^2 - \left\Vert \bm x- \frac{\bm \alpha\transp \bm x}{\Vert \bm\alpha \Vert^2} \bm \alpha \right\Vert^2 \right)} \notag\\
 & =\e^{\frac{|\bm \alpha\transp \bm x|^2}{\Vert \bm \alpha\Vert^2  P_w}} \gtrless \gamma'
\end{align}
i.e.,
\begin{equation}
 \frac{|\bm \alpha\transp \bm x|^2}{\Vert \bm \alpha\Vert^2  P_w} = \frac{|\alpha x_1+ \alpha_{sr}x_2|^2}{(\alpha^2 +\alpha_{sr}^2)  P_w}  \gtrless \ln \gamma' =\gamma 
\end{equation}
as shown is Eq.~(6a). The other cases can similarly be handled.

\subsection{SNRs in Eqs.~(7)}

The test statistic in Eq.~(6), Case~a, requires computing the quantity $\alpha x_1+ \alpha_{sr}x_2=\bm \alpha\transp \bm x$. From Eq.~\eqref{obs_model} we have
\begin{equation}
 \bm \alpha\transp \bm x = \bm \alpha\transp (A\bm \alpha + \bm w) = A \Vert \bm \alpha \Vert^2 + \bm \alpha\transp \bm w
\end{equation}
so that
\begin{equation}
 \SNR_a= \frac{\mathbb E[|A|]\Vert \bm \alpha\Vert^4}{\Vert \bm \alpha \Vert^2 P_w} =  \frac{\mathbb E[\sigma]\Vert \bm \alpha\Vert^2}{P_w} =\frac{\bar \sigma(\alpha^2+ \alpha_{sr}^2)}{P_w}
\end{equation}
as reported in Eq.~(7a). The other cases can similarly be handled. Finally, as to the maximization of $\SNR_b$ and $\SNR_c$ in Eqs.~(7b) and~(7c), we have
\begin{align}
 \max_{\epsilon\in[0,1]} \SNR_b &=  \max_{\epsilon\in[0,1]} \frac{\bar\sigma \alpha^2 \epsilon +\alpha_{st}^2 (1-\epsilon)}{P_w}\notag\\
 & = \frac{\bar\sigma \max \{\alpha^2 , \alpha_{st}^2\}}{P_w}\notag\\
 & = \frac{\bar\sigma \alpha^2}{P_w} \max \left\{1 , \frac{\alpha_{st}^2}{\alpha^2} \right\} \notag\\
 &= \SNR_0 \max\{1, K_{st}\}
\end{align}
for $\epsilon = \epsilon^*_b =\mathbbm 1_{\{ \alpha^2_{st}\leq  \alpha^2\}} =\mathbbm 1_{\{K_{st}\leq 1\}}$, and
\begin{align}
 \max_{\epsilon\in[0,1]} \SNR_b &=\max_{\epsilon\in[0,1]} \frac{\bar\sigma (\alpha \sqrt{\epsilon} +\alpha_{st} \sqrt{1-\epsilon})^2}{P_w}\notag\\
 & = \frac{\bar\sigma \alpha^2 +\alpha_{st}^2}{P_w} \\
 & = \frac{\bar\sigma \alpha^2}{P_w}  \left(1 + \frac{\alpha_{st}^2}{\alpha^2}\right) \notag\\
 & = \SNR_0 (1+ K_{st})
\end{align}
for $\epsilon = \epsilon_c^*= \frac{\alpha^2}{\alpha^2+\alpha_{st}^2} = \frac{1}{1+K_{st}}$, respectively.

\subsection{Detection probability in the closely-spaced scenario}

Assuming\footnote{These models are known as Marcum's non fluctuating case, Swerling's case~1 or~3 (scan-to-scan or pulse-to-pulse fluctuation), and Swerling's case~2 or~4 (scan-to-scan or pulse-to-pulse fluctuation), respectively.} $\sigma=\bar \sigma$ non fluctuating, $\sigma$ exponentially distributed with mean $\bar \sigma$, or $\sigma$ gamma distributed with mean $\bar \sigma$ and variance $\bar \sigma^2/2$, we have that the detection probability is
\begin{equation}
 P_\text{d}=\begin{cases}
Q\big(\sqrt{2\SNR},\sqrt{2\gamma}\big) & \text{(non fluctuating)}\\
 \e^{-\frac{\gamma}{1+\SNR}} & \text{(exponential)}\\
 \left(1+ \frac{\gamma \frac{\SNR}{2}}{\left(1+\frac{\SNR}{2}\right)^2} \right) \e^{-\frac{\gamma}{1+\frac{\SNR}{2}}} & \text{(gamma)}
 \end{cases}
\end{equation}
respectively, where $\SNR$ is as in Eqs.~(7), and $Q(\,\cdot\,,\,\cdot\,)$ is the Marcum $Q$-function~[16].

\subsection{RIS-target channel phases in Footnote~5}

\begin{figure}[t]
 \centering
 \centerline{\includegraphics[width=1.05\columnwidth]{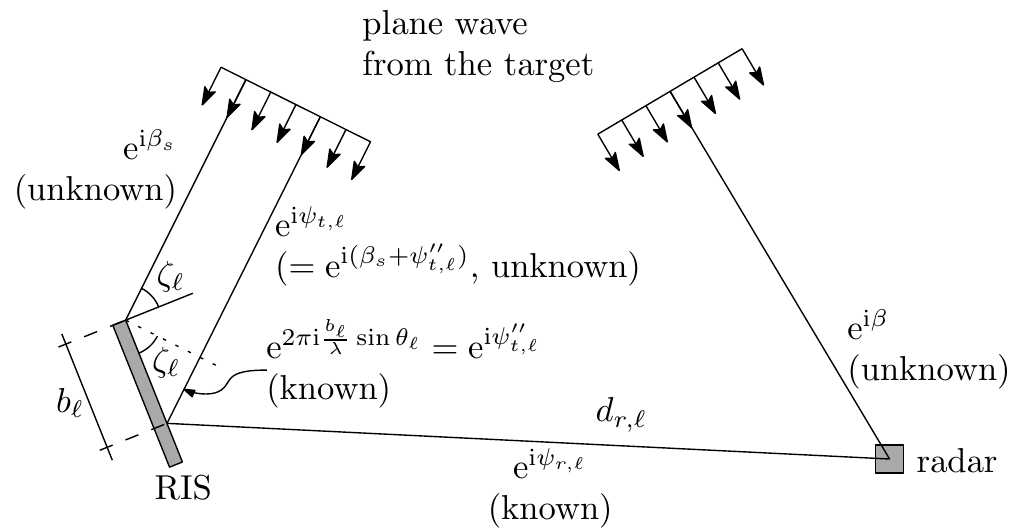}}
\caption{Phases of the RIS-target channel in the widely-spaced scenario.} \label{fig_5}
\end{figure}

Radar and RIS \emph{see} the target from different aspect angles, and, since a plane wave is impinging on the RIS, the phases of the target-RIS channel can be expressed as $\psi_{t,\ell}''+\beta_s$, where $\beta_s$ is the phase of the channel between the target and the first element of the RIS, and $\psi_{t,\ell}''$ is known and depends only on the mutual position of the first and $\ell$-th element of the RIS with respect to the target; $\{\psi_{t,\ell}''\}_{\ell=1}^L$ are in fact the phases of a steering vector, as it can also be seen from Fig.~\ref{fig_5}. In this case, we have two different RCSs of the target, $\sqrt{\sigma} \e^{\i\beta}$ and $\sqrt{\sigma_s} \e^{\i\beta_s}$, in the two echoes, and they can be modeled as independent random variables.

\subsection{GLRT's in Eqs.~(11) and SNRs in Eqs.~(12)}

They can be proved as in the previous sections.

\subsection{SNRs in Eqs.~(13) and (14)}

From Eq.~(10c), the SNR in Case~c is
\begin{align}
 \SNR_c &= \frac{1}{P_w}\mathbb E\left[ \left| \sqrt{\sigma\epsilon}\e^{\i \beta} \alpha +  \sqrt{\sigma_s (1-\epsilon)}\e^{\i \beta_s} \alpha_{st} \right|^2\right] \notag\\
 &= \frac{1}{P_w} \Big( \mathbb E[\sigma] \alpha^2\epsilon + 2\Re\left\{ \mathbb E\left[ \sqrt{\sigma} \e^{\i\beta} \sqrt{\sigma_s} \e^{-\i\beta_s}  \right] \right\} \notag\\
 & \quad \times \sqrt{\epsilon(1-\epsilon)} \alpha \alpha_{st} + \mathbb E[\sigma_s] \alpha_{st}^2 (1-\epsilon) \Big).
\end{align}
Since the two target responses, $\sqrt{\sigma} \e^{\i\beta}$ and $\sqrt{\sigma_s} \e^{\i\beta_s}$ are independent, and the phases $\beta$ and $\beta_s$ are uniformly distributed over $[0,2\pi)$, we have
\begin{equation}
  \mathbb E\bigl[ \sqrt{\sigma} \e^{\i\beta} \sqrt{\sigma_s} \e^{-\i\beta_s} \bigr]  =  \mathbb E\left[ \sqrt{\sigma} \e^{\i\beta} \right]  \mathbb E\left[ \sqrt{\sigma_s} \e^{-\i\beta_s}  \right] =0
\end{equation}
and, therefore,
\begin{align}
 \SNR_c &= \frac{\epsilon  \bar \sigma \alpha^2 + (1-\epsilon) \bar \sigma_s \alpha_{st}^2}{P_w} 
\end{align}
as shown in Eq.~(13). Concerning its optimization over $\epsilon \in[0,1]$, if $\bar \sigma\in [\bar \sigma_\text{min}, \bar \sigma_\text{max}]$ and $\bar \sigma_s\in[\bar \sigma_{s,\text{min}}, \bar \sigma_{s,\text{max}}]$, then the maximization of the worst-case SNR gives
\begin{align}
 \epsilon^*_c&=  \argmax_{\epsilon \in [0,1]} \min_{\substack{\bar \sigma\in [\bar \sigma_\text{min}, \bar \sigma_\text{max}] \\ \bar \sigma_s\in [\bar \sigma_\text{s,min}, \bar \sigma_\text{s,max}]}}  \frac{\epsilon  \bar \sigma \alpha^2 + (1-\epsilon) \bar \sigma_s \alpha_{st}^2}{P_w} \notag\\
 &= \argmax_{\epsilon \in [0,1]} \frac{\epsilon  \bar \sigma_\text{min} \alpha^2 + (1-\epsilon) \bar \sigma_{s,\text{min}} \alpha_{st}^2}{P_w} \notag\\
 & =\mathbbm 1_{\{ \bar\sigma_{s,\text{min}}\alpha^2_{st} \leq \bar\sigma_\text{min}\alpha^2\}} \notag\\
 &=\mathbbm 1_{\{K_{st} \bar\sigma_{s,\text{min}} / \bar\sigma_\text{min} \leq 1\}}
\end{align}
and this choice results in
\begin{align}
 \SNR_c \big|_{\epsilon=\epsilon_c^*} &=  \frac{\epsilon  \bar \sigma \alpha^2 + (1-\epsilon) \bar \sigma_s \alpha_{st}^2}{P_w} \Bigg|_{\epsilon=\epsilon_c^*}\notag\\
 &= \begin{cases} \frac{\bar\sigma \alpha^2}{P_w} = \SNR_0, \text{ if } K_{st} \bar\sigma_{s,\text{min}} / \bar\sigma_\text{min} \leq 1\\ \frac{\bar\sigma_s \alpha_{st}^2}{P_w} = \frac{\alpha_{st}^2}{\alpha^2} \frac{\bar\sigma_s}{\bar \sigma} \frac{\bar\sigma \alpha^2}{P_w} = K_{st} \frac{\bar\sigma_s}{\bar \sigma} \SNR_0 ,\text{ otherwise} \end{cases}\notag\\
 &=\SNR_0 \Bigl(\mathbbm 1_{\{K_{st} \bar \sigma_{s,\text{min}} / \bar \sigma_\text{min} \leq 1  \}} \notag\\
 &\quad +\mathbbm 1_{\{K_{st} \bar \sigma_{s,\text{min}} / \bar \sigma_\text{min} \geq 1 \}} K_{st} \frac{\bar \sigma_s}{\bar \sigma} \bigr)
\end{align}
as in Eq.~(14).

\subsection{Detection probability in the widely-spaced scenario}

In Case~c, the probability of detection is
\begin{equation}
 P_\text{d}=\begin{cases}
Q\big(\sqrt{2\SNR_c},\sqrt{2\gamma}\big) & \text{(non fluctuating)}\\
 \e^{-\frac{\gamma}{1+\SNR_c}} & \text{(exponential)}\\
 \left(1+ \frac{\gamma \frac{\SNR_c}{2}}{\left(1+\frac{\SNR_c}{2}\right)^2} \right) \e^{-\frac{\gamma}{1+\frac{\SNR_c}{2}}} & \text{(gamma)}
 \end{cases}
\end{equation}
respectively, where $\SNR$ is as in Eqs.~(7). As to Cases~a and~b, $P_\text{d}$ does not admit a simple expression for these fluctuation models, since the SNRs of the two observations are in general different; for the exponential distribution, however, we have
\begin{equation}
P_\text{d}= \frac{1+\SNR_2}{\SNR_1-\SNR_2} \e^{-\frac{\gamma}{1+\SNR_1}} - \frac{1+\SNR_1}{\SNR_1-\SNR_2} \e^{-\frac{\gamma}{1+\SNR_2}}
\end{equation}
where $\SNR_1$ and $\SNR_2$ are those in Eq.~(12).

\subsection{RIS gains in Eq.~(15)}

The approximations in Sec.~V are
\begin{subequations}
\begin{align}
 (\theta_{r,\ell}, \omega_{r,\ell})& \approx (\theta_r, \omega_r)\\
 S_{sr,\ell} &\approx \pi (\lambda/2)^2 \cos\theta_t \cos \omega_t \cos\theta_r \cos \omega_r \notag\\
 &=\frac{1}{L^2} \underbrace{L(\lambda/2)^2  \cos\theta_r \cos \omega_r}_{A_{sr}} \underbrace{L \pi \cos\theta_t \cos \omega_t}_{G_{st}}\notag\\
 &=\frac{A_{sr} G_{st}}{L^2}\\
 d_{r,\ell}& \approx d_r\\
 G_{rs,\ell}&\approx \frac{4\pi d_r^2}{A_{rs}} \mathbbm{1}_{\{\ell\in\mathcal B\}}
\end{align}%
\end{subequations}
where $A_{sr}$ is the effective area of the RIS seen from the radar, $G_{st}$ is the gain of the RIS (seen as an aperture antenna) towards the target, $A_{rs}$ is the surface area covered by the 3-dB beamwidth of the radar antenna at a distance equal to $d_r$, and $\mathcal B$ is the set containing the indexes of the elements of the RIS that fall in the 3-dB beamwidth of the radar. Therefore, the gain $K_{sr}$ in Eq.~(8) can be approximated as
\begin{align}
 K_{sr} & \approx \frac{\rho^2}{4\pi d_t^2 G_{rt}} \left( \sum_{\ell=1}^L \sqrt{\frac{4\pi}{A_{rs}} \mathbbm{1}_{\{\ell\in\mathcal B\}} \frac{A_{sr} G_{st}}{L^2}} \right)^2\notag\\
 &=  \frac{\rho^2 G_{st}}{d_t^2 G_{rt}} \frac{A_{sr}}{A_{rs}} \frac{1}{L^2} \left( \sum_{\ell=1}^L \mathbbm{1}_{\{\ell\in\mathcal B\}} \right)^2\notag\\
 &\approx \begin{cases} \frac{\rho^2 G_{st}}{d_t^2 G_{rt}} \frac{A_{sr}}{A_{rs}} \frac{1}{L^2} L^2, & \text{if }A_{rs}\geq A_{sr}\\ \frac{\rho^2 G_{st}}{d_t^2 G_{rt}} \frac{A_{sr}}{A_{rs}} \frac{1}{L^2} \left(\frac{A_{rs}}{A_{sr}}\right)^2, & \text{otherwise} \end{cases} \notag\\
 &=\frac{\rho^2 G_{st}}{d_t^2 G_{rt}} \min\left\{ \frac{A_{sr}}{A_{rs}} , \frac{A_{rs}}{A_{sr}}  \right\}
\end{align}
as in Eq.~(15), where, in the third line, we have exploited the fact that $\sum_{\ell=1}^L \mathbbm{1}_{\{\ell\in\mathcal B\}}=L$, if the 3-dB beamwidth of the radar covers the entire RIS, i.e., if $A_{rs}\geq A_{sr}$, and $\sum_{\ell=1}^L \mathbbm{1}_{\{\ell\in\mathcal B\}}\approx A_{rs}/ A_{sr}$, otherwise. The same approximation holds for $K_{st}$, since $S_{st,\ell}=S_{sr,\ell}$.

\subsection{Radar Equation in~(16)}

Under the assumption that $A_{rs}\approx A_{sr}$ (i.e., that the area covered by the 3-dB beamwidth of the radar equals the effective area of the RIS seen from the radar), the SNR of the indirect echo in Cases~b and~c is, from Eqs.~(3) and~(15),
\begin{align}
\frac{\bar \sigma \alpha_{st}^2}{P_w} &= \frac{\bar \sigma \alpha^2}{P_w} \frac{\alpha_{st}^2}{\alpha^2} \notag\\
&= \frac{P_r G_{rt}^2\lambda^2 \bar\sigma}{(4\pi)^3 \rho^4 P_w}  K_{st}\notag\\
& \approx \frac{P_r G_{rt}^2\lambda^2 \bar\sigma}{(4\pi)^3 \rho^4 P_w}  \frac{\rho^2 G_{st}}{d_t^2 G_{rt}} \notag\\
&= \frac{P_r G_{rt} G_{st} \lambda^2 \bar\sigma}{(4\pi)^3 \rho^2 d_t^2 P_w}\notag
\end{align}
as Eq.~(16). The same approximation holds for the SNR of the indirect echo in Case~a, $\bar \sigma \alpha_{sr}^2/P_w$, since $K_{sr}=K_{st}$.

\bibliographystyle{IEEEtran}

\end{document}